\documentstyle[aps,epsfig,twocolumn,floats]{revtex}

\begin{document}                                                                

\title{
Modified Empirical Parametrization
of Fragmentation Cross Sections
}

\author{
{\sc K.~S\"ummerer$^1${\footnotemark[1]}}
and {\sc B.~Blank$^2${\footnotemark[3]}}
}

\address{                                                               
$^1$Gesellschaft f\"ur Schwerionenforschung, 
Planckstr.1, D-64291 Darmstadt, Germany\\
$^2$CEN Bordeaux-Gradignan, Le Haut-Vigneau, 
F-33175 Gradignan Cedex, France
}
\date{October 20, 1999}

\maketitle
                                                                                
\begin{abstract}                                                                
New experimental data obtained mainly at the GSI/FRS facility allow
to modify the empirical parametrization
of fragmentation cross sections, {\sc EPAX}. 
It will be shown that minor
modifications of the parameters lead to a much better reproduction
of measured cross sections.
The most significant changes refer to the description of
fragmentation yields close to the projectile and
of the memory effect of neutron-deficient projectiles.
\end{abstract}

\pacs{PACS: 25.70.Mn}

\renewcommand{\thefootnote}{\fnsymbol{footnote}}
\footnotetext[1]{Electronic address: k.suemmerer@gsi.de}
\footnotetext[3]{Electronic address: blank@cenbg.in2p3.fr} 
                                                                              
\section{Introduction}
The pioneering experiments of projectile
fragmentation at relativistic energies of $^{40}$Ar and $^{48}$Ca beams 
at the LBL Bevalac~\cite{viyogi,ca48}
have demonstrated the potential of this method for the production
of exotic nuclei.
Based on these ideas,
the SIS/FRS facility~\cite{frs} at 
GSI has used also heavier projectiles like e.g. $^{58}$Ni,
$^{86}$Kr, $^{129}$Xe, and $^{208}$Pb
to produce and study exotic nuclei~\cite{blank,weber,joerg,dejong}. 
For planning such experiments, 
when count-rate predictions are needed,
analytical descriptions of fragmentation cross sections are useful.
They are also useful in 
simulation programs for projectile-fragment separators
(like e.g. {\sc INTENSITY}~\cite{intensity} or {\sc MOCADI}~\cite{mocadi}).
Compared to physical models
of high-energy fragmentation reactions, 
which in general involve time-consuming
Monte-Carlo calculations, the virtue of an analytical formula
lies in the short computing time and the possibility to calculate
easily sub-microbarn cross sections that are beyond the reach of 
physical-model calculations.

In 1990, S\"ummerer {\it et al.}~\cite{epaxv11} proposed
a universal Empirical PArametrization 
of fragmentation CROSS sections (``{\sc EPAX}'', Ref.~\cite{epaxv11})
which was based on and similar to previous prescriptions by
Rudstam~\cite{rudstam} and Silberberg {\it et al.}~\cite{st1}.
The parametrization was to a large extent based on multi-GeV proton-induced 
spallation cross sections, since only scarce heavy-ion induced 
experimental data were available at that time.
Meanwhile, more precise data from relativistic heavy-ion-induced
fragmentation reactions 
together with recent results from projectile fragmentation
of heavy nuclei ($^{197}$Au and $^{208}$Pb)
on H$_2$ targets~\cite{farget,enqvist}) allow a more stringent
comparison of proton- and heavy-ion induced isotope distributions.
This comparison indicates that for heavy nuclei the two
reactions lead to different isotopic distributions,
which cannot be obtained from each other just by
scaling factors.
This can be easily understood since 
heavy-ion induced reactions are expected to
deposit more excitation energy in a nucleus than proton-induced
reactions, making the final product distributions
--- after evaporation --- 
broader and more neutron-deficient.
Nevertheless, the data show that in both cases
the isotopic yield distributions can 
be well described by Gaussian-like
analytical functions with parameters that vary smoothly as a function
of fragment mass~\cite{epaxv11}. 
In the present paper, we will base the choice of these parameters
exclusively on heavy-ion-induced reaction data.  
  
We will first review briefly the basic characteristics of the {\sc EPAX}
formula and then show which modifications are necessary to improve
the accuracy with which the new experimental results can be
reproduced. This will be followed by a brief comparison with
similar attempts by other authors.

\section{The {\sc EPAX} formula}

\subsection{Basic characteristics}

The basic characteristics of the analytical description of
high-energy fragmentation cross sections by the {\sc EPAX} formula are the
following~\cite{epaxv11}:

\begin{itemize}

\item In the absence of systematic excitation-function measurements
of heavy-ion induced fragmentation reactions, the
formula is valid only for the so-called "limiting fragmentation"
regime, i.e. for projectile energies where the fragmentation yields
are no longer energy dependent, at least within the accuracy of
the formula (approximately within a factor of 2). 
This is certainly true for incident energies
considerably above the Fermi energy in nuclei 
($\approx$ 40 $A$ MeV), in particular for the
typical SIS energies of 500 to 1000 $A$ MeV.

\item The EPAX formula is meant to describe the fragmentation
of medium- to heavy-mass projectiles;
nucleon-pickup cross sections are not included.
No attempt is made to describe the fragmentation of fissile nuclei.
Therefore, the range of validity is limited to projectiles 
from around argon to below
the lead and bismuth isotopes. Predictions for production cross sections
of fission products or of fragments below U where fission
competition is significant
require an elaborate description of
the fission process, such as can be found e.g. in a recent publication
by Benlliure {\it et al.}~\cite{jose}.

\item For fragments sufficiently far away from the projectile (i.e. for
mass losses larger than 15-20\% of the projectile mass),
the isotope distributions are largely independent of the original nucleus;
their position, shape, and width depend only on the fragment mass number.
This constitutes what has been termed the ``residue corridor'' 
and is related to the fact that the isotope distributions are mainly
governed by statistical evaporation from highly excited prefragments
produced in collisions between relativistic heavy ions.

\item For fragments that involve only a small mass loss from the projectile,
the isotope distributions should be centered close to the projectile
and their variance should be small. Therefore, a smooth transition is
anticipated between the residue corridor and the projectile. 
The parametrization of this smooth transition constitutes the main task
in designing the formula.

\end{itemize}

In a first step, a parameter set has been searched for that describes the
fragmentation yields from projectiles located close to the line
of $\beta-$stability. In a second step, a modification of the yield
distributions due to the neutron or proton excess of projectiles
located on the neutron- or proton-rich side of the line of $\beta-$stability 
(the "memory effect") has been parametrized.

\subsection{The parameters of {\sc EPAX} Version 1}

As explained in detail in Ref.~\cite{epaxv11}, the cross section (in barn)
of a fragment with mass $A$ and charge $Z$ produced by projectile
fragmentation from a projectile $(A_p,Z_p)$ impinging
on a target $(A_t,Z_t)$ is written as
\begin{eqnarray}                             
\sigma (A,Z)& = &Y_A~\cdot~\sigma (Z_{prob} - Z)\nonumber\\
            & = &Y_A~\cdot~n~\cdot 
		exp(-R\cdot |Z_{prob} - Z|^{U_{n(p)}})\label{eq1}
\end{eqnarray}    
The first term, $Y_A$, represents the mass yield, 
i.e. the sum of the isobaric  
cross sections with fragment mass $A$.  
The second term describes the "charge dispersion", 
the distribution of elemental cross sections with a given mass around its       
maximum, $Z_{prob}$. 
The shape of the charge dispersion is controlled by the width   
parameter, $R$, and the exponent, $U_n$($U_p$),
on the neutron-(proton-) rich
side of the residue corridor.
The factor $n = \sqrt{R/\pi}$ simply serves to normalize
the integral of the charge dispersion to unity.

The mass-yield curve is taken to be an exponential as a function of
$A_p - A$. The slope of this exponential, $P$, is a function of
the projectile mass. An overall scaling factor, $S$, accounts for the
peripheral nature of fragmentation reactions
and therefore depends on the circumference of the colliding
nuclei:
\begin{eqnarray}                                                                
Y_A & = & S   \cdot P \cdot exp(-P \cdot (A_p-A))\label{eq2} \\         
S   & = & S_2 \cdot (A_p^{1/3} + A_t^{1/3} + S_1)\;\; [{\rm barn}]\label{eq3} \\
ln~P & = & P_2 \cdot A_p + P_1 \label{eq4} 
\end{eqnarray}
\noindent
The numerical values of the various constants can be found in 
Table~\ref{tab_const}.
\narrowtext
\begin{table}[bh]
\caption{Constants used in EPAX Version 1 
(Ref.\protect\cite{epaxv11}) 
and those used in Version 2 (this work). Note that in a few cases also
the functional form to calculate a parameter has changed (see text).}
\begin{tabular}{ccll}
Parameter & Constant & \multicolumn{2}{c}{Value} \\
          &          & Version 1  & Version 2 \\
\tableline
Scaling factor & $S_1$ & -2.38            & -2.38 \\
  $S$          & $S_2$ & 0.450            &  0.270 \\
\tableline
mass yield slope & $P_1$ & -2.584            & -2.584   \\
  $P$            & $P_2$ & $-7.57\cdot 10^{-3}$   & $-7.57\cdot 10^{-3}$  \\
\tableline
width parameter  & $R_1$ & 0.778             & 0.885   \\
  $R$            & $R_2$ & $-6.77\cdot 10^{-3}$ & $-9.82\cdot 10^{-3}$  \\
\tableline
$Z_{prob}$ shift  & $\Delta_1$ & 0.895           &  -1.09  \\
  $\Delta$         & $\Delta_2$ & $2.70\cdot 10^{-2}$ 
			&  $3.05\cdot 10^{-2}$  \\
                   & $\Delta_3$ & $2.04\cdot 10^{-4}$ 
			&  $2.14\cdot 10^{-4}$  \\
                   & $\Delta_4$ & 66.22           &  71.35  \\
\tableline
n-rich slope       & $U_n$ & 1.50           & 1.65       \\
  $U_n$            &       &                &             \\
\tableline
p-rich slope       & $U_1$ & 2.00           & 1.79  \\
  $U_p$            & $U_2$ &                & $ 4.72\cdot 10^{-3}$ \\
                   & $U_3$ &                & $-1.30\cdot 10^{-5}$ \\
\tableline
n-rich memory effect & $n_1 $ & 0.40            & 0.40  \\
  $\Delta_m$         & $n_2 $ & 0.60            & 0.60  \\
\tableline
p-rich memory effect & $p_1 $ & 0.00            & -10.25 \\
  $\Delta_m$         & $p_2 $ & 0.60            &  10.10 \\
\tableline
correction factor & $d_1 $ & -51.0          & -25.0 \\
  for $\Delta$    & $d_2 $ & 0.86           & 0.80  \\
\tableline
correction factor & $r_1 $ & 20.0           & 20.0  \\
  for $R$         & $r_2 $ & 0.86           & 0.82 \\
\tableline
correction factor   & $y_1 $ &                & 200.0 \\
  for $Y_A$         & $y_2 $ &                & 0.90  \\
\end{tabular}
\label{tab_const}
\end{table}
The charge dispersion is characterized
by the three parameters $R$, $Z_{prob}$, and $U$.
These three parameters are                                                      
strongly correlated and difficult to obtain uniquely with a
least-squares fitting technique.
Note that the isobar distributions are not symmetric on the 
neutron- and proton-rich side, therefore $U$ has two different values,
$U_p$ and $U_n$, on the proton- or neutron-rich side of the
valley of $\beta$-stability, respectively.
In Ref.~\cite{epaxv11}, the exponent $U$ for the neutron-rich side of
the isobar distribution was chosen as
$U_n=1.5$, whereas the proton-rich side falls off
like a Gaussian ($U_p=2$).
The maxima of the charge dispersions, $Z_{prob}$, have been
parametrized relative to the valley of $\beta$-stability,
\begin{equation}                                                                
Z_{prob} = Z_{\beta} + \Delta                                                  
\label{eq5}                                                                    
\end{equation}

$Z_{\beta}$ is approximated by the smooth function
\begin{equation}                                                                
Z_{\beta}=A/(1.98 + 0.0155 \cdot A^{2/3})                                    
\label{eq6}                                                                     
\end{equation}                                                                  
                                                                                
$\Delta$ is found to be a linear function of
the fragment mass, $A$, for heavy fragments ($A\geq\Delta_4$), 
and is extrapolated quadratically to zero:
\begin{equation}                                                                
\Delta=\left\{ \begin{array}{ll}                                                
        \Delta_3 \cdot A^{2}                                           
        &\;\;\mbox{if $A < \Delta_4$} \\
        \Delta_2 \cdot A + \Delta_1  &
          \;\;\mbox{if $A \geq \Delta_4$} \end{array}
        \right.\label{eq7}
\end{equation}
Similar to the parameter $Z_{prob}$ just discussed, the
width parameter, $R$, is a function of fragment mass only, irrespective of
the projectile.
In Ref.~\cite{epaxv11} it was found that the experimental $R$-values
can be approximated by an exponential of the form
\begin{equation}
ln~R = R_2 \cdot A + R_1
\label{eq8}    
\end{equation}

The equations given above are sufficient to describe the "residue corridor",
i.e. the yield distributions of projectiles located on
the line of $\beta$-stability if the fragment mass is far from the
projectile mass. Close to the projectile, the following modifications
have to be introduced~\cite{epaxv11}:
\begin{eqnarray}
\Delta & = & \Delta \cdot [1+d_1 \cdot (A/A_p-d_2)^2] \label{eq9}\\
R & = & R \cdot [1+r_1 \cdot (A/A_p-r_2)^2] \label{eq10}
\end{eqnarray}
This serves to gradually reduce to zero the offset of $Z_{prob}$ from
the line of $\beta$-stability and to decrease the width of the
charge dispersion when $A$ approaches the mass of the projectile, $A_p$.

A final correction applies if the projectile does not lie on the
line of $\beta$-stability. 
In this case, the $A/Z$ ratio of the fragments will to some extent
``remember'' the $A_p/Z_p$ ratio of the projectile
(``memory effect'').
For an analytical description, it is simply assumed that
the charge dispersions are shifted by an amount $\Delta_m$
which is a certain fraction               
of the distance of $Z_p$ from $Z_{\beta p}$,
the nuclear charge on the line of $\beta$-stability for $A_p$.
Close to the projectile, this fraction is clearly close to unity
(full memory effect),
whereas it should gradually approach zero with increasing
distance of the fragment mass $A$ from the projectile (loss of memory). 
The shape of the isobar distribution is assumed to be unchanged.
Thus 
\begin{equation}
Z_{prob} = Z_{\beta} + \Delta + \Delta_m
\label{eq11}                                                                   
\end{equation}                                                                  
where $\Delta_m$ for neutron-rich projectiles is given by
\begin{equation}
\Delta_m=[n_1\cdot (A/A_t)^2+n_2\cdot (A/A_t)^4]\cdot (Z_p-Z_{\beta p})
\label{eq12}
\end{equation}
The corresponding constants for proton-rich projectiles are termed
$p_1$ and $p_2$ instead of $n_1$ and $n_2$. 
Numerical values for all constants are given in Table \ref{tab_const}.

\begin{figure}[tbh]
\begin{center}
\epsfig{file=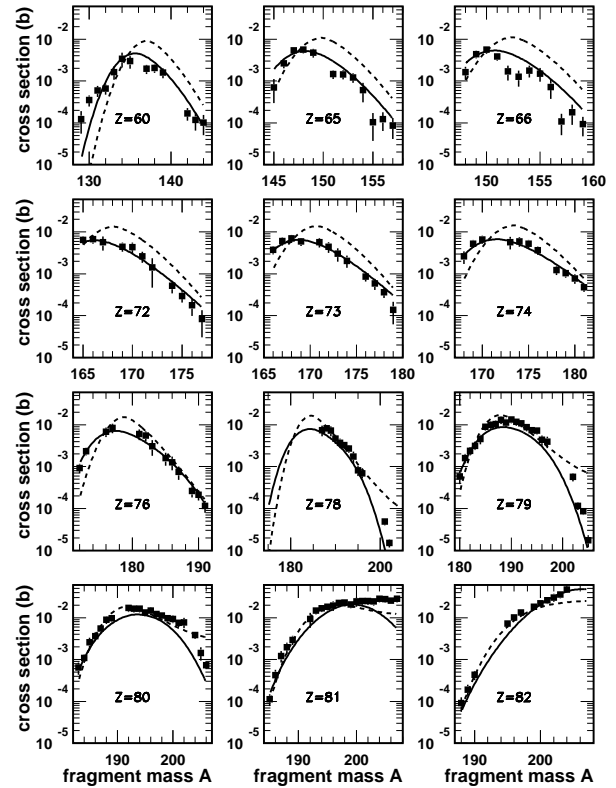,width=1.0\linewidth}
\end{center}
\caption{
Isotope distributions from 1 $A$ GeV $^{208}$Pb 
fragmentation in a $^{nat}$Cu target~\protect\cite{dejong}
in comparison with the old EPAX 
parametrization Version 1 (dashed curves) and the modified Version~2 
(full curves).
} 
\label{fig_pb}
\end{figure}

With this parametrization, {\sc EPAX} Version 1 was rather successful
in describing the gross features of isotope distributions of
high-energy projectile fragmentation.
This has been visualized e.g. in
Refs.~\cite{blank,weber,joerg,dejong},
where comparisons of {\sc EPAX} with experimental data over many orders
of magnitude in cross section can be found.
As a particular example, we plot in Fig.\ref{fig_pb} experimental 
isotope distributions from 
1 $A$ GeV $^{208}$Pb fragmentation~\cite{dejong} 
in comparison with the {\sc EPAX} Version 1 parametrization
(dashed curves).

This set of data was also chosen to illustrate that the old {\sc EPAX} version 
has problems to reproduce satisfactorily the isotope distributions
of very heavy fragments. As can be seen best for the low-$Z$
isotope distributions, the dashed lines are centered too much on the
neutron-rich side and exhibit too large maxima.
Moreover, the cross sections for fragments close to the projectile
(with masses $A \ge$ 200) are predicted much too high.
Minor discrepancies were also found for $^{129}$Xe-fragment yields
~\cite{joerg} where the experimental distributions were found to be 
wider than the {\sc EPAX} predictions. Serious deficiencies were 
revealed when, compared to the {\sc EPAX} prediction, too large 
cross sections for neutron-deficient isotopes
from $^{58}$Ni fragmentation~\cite{blank} and too small cross sections of
neutron-deficient Sn isotopes from
$^{112}$Sn fragmentation~\cite{sn112} were measured.

\section{Modifications of the {\sc EPAX} formula}

\subsection{Fragmentation of projectiles close to $\beta$-stability}

The availability of extensive cross-section measurements for
projectile fragmentation of $^{40}$Ar, $^{58}$Ni,
$^{86}$Kr, $^{129}$Xe, and $^{208}$Pb~\cite{aki,blank,weber,joerg,dejong}
down to $\mu$b or nb cross sections allowed in a first step to
find a better parametrization of the mass yields $Y(A)$. 
In a second step, we adjusted the parameters describing
the residue corridor, i.e.
the width parameter, $R$, and
the slope constants, $U_n$ and $U_p$, of the quasi-Gaussian charge
dispersion together with its centroid, $Z_{prob}$.
In a third step, the correction factors
for isotopic yields close to the projectile were modified.
Finally, modifications for projectiles outside the
valley of $\beta$-stability were redetermined.
Numerical values for
the new constants are given in the fourth column of Table \ref{tab_const}.

\subsubsection{Integrated mass and charge yields}

\begin{figure}[bth]
\begin{center}
\epsfig{file=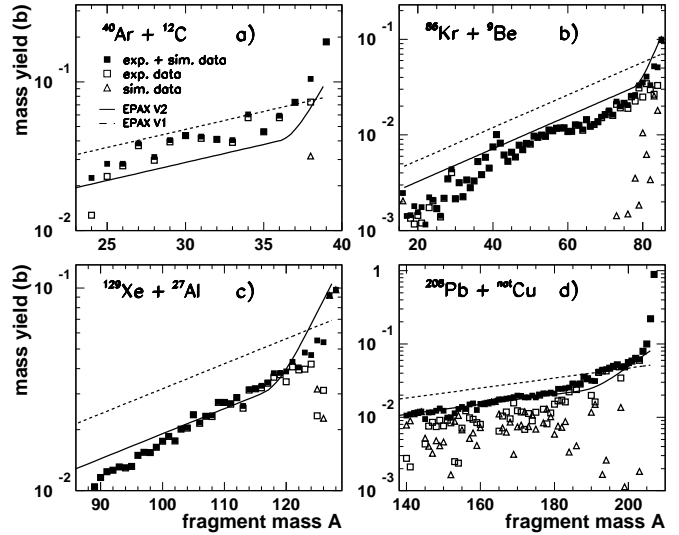,width=1.10\linewidth}
\end{center}
\caption{Experimental mass yields, $Y(A)$, for the reactions of
         $^{40}$Ar+$^{12}$C (a, Ref.~\protect\cite{webber1}),
         $^{86}$Kr+$^{9}$Be (b, Ref.~\protect\cite{weber}), 
	 $^{129}$Xe+$^{27}$Al (c, Ref.~\protect\cite{joerg}), and 
	 $^{208}$Pb+$^{nat}$Cu (d, Ref.~\protect\cite{dejong}) compared 
         to the old EPAX parametrization Version 1 (dashed line) and to the
         new EPAX formula Version 2 (solid line). 
         The open squares are the sum of
         the experimental cross sections for a given fragment mass $A$, whereas 
         the open triangles represent the calculated cross sections using the 
         new EPAX formula for isotopes missing in the experimental data
         for a given mass $A$. The full squares are the sum of the two
         values.
} 
\label{a_yield}
\end{figure}

According to Eqs.~(\ref{eq2}-\ref{eq4}), $Y(A)$ is described by 
an exponential function depending on the fragment mass, $A$, the projectile 
mass, $A_p$, and the target mass, $A_t$. In {\sc EPAX} Version 1, no additional 
correction close to the projectile has been performed. As shown 
by the dashed lines in 
Fig.~\ref{a_yield}, this yields only a very rough agreement with experimental 
data.
Note that the experimental mass yields had to be complemented
by calculated cross sections (from {\sc EPAX} Version 2) where experimental
data points were missing.  For the $^{40}$Ar, $^{86}$Kr, and $^{129}$Xe data,
the additional calculated cross sections 
contribute only close to the projectile; in the $^{208}$Pb case, however,
the experimental data are less complete and more significant
corrections were necessary.
 
Fig.~\ref{a_yield} shows that the slope somewhat further away 
from the projectile is
reasonably well reproduced, however, 
the absolute height as well as the slope close to
the projectile mass needed to be modified.
A much better overall normalisation was achieved by reducing
the scaling factor $S_2$ by a factor of 0.6, as demonstrated 
by the solid lines in Fig.~\ref{a_yield}.
An additional correction introduced in the new {\sc EPAX} version
consists of an increase in the mass yield, $Y_A$, 
(Eq.~\ref{eq2}) close to the projectile according to
\begin{equation}
Y_A  = Y_A \cdot [1+y_1 \cdot (A/A_p-y_2)^2]. \label{eq15}
\end{equation}
for $A/A_p \ge y_2$.
These modifications give a satisfactory agreement with experimental data. 
In the case of $^{129}$Xe, the slope far from the 
projectile mass deserves further attention. However, due to the lack of 
other complete experimental data, we were not able to significantly improve
the mass-yield description.

\begin{figure}[bth]
\begin{center}
\epsfig{file=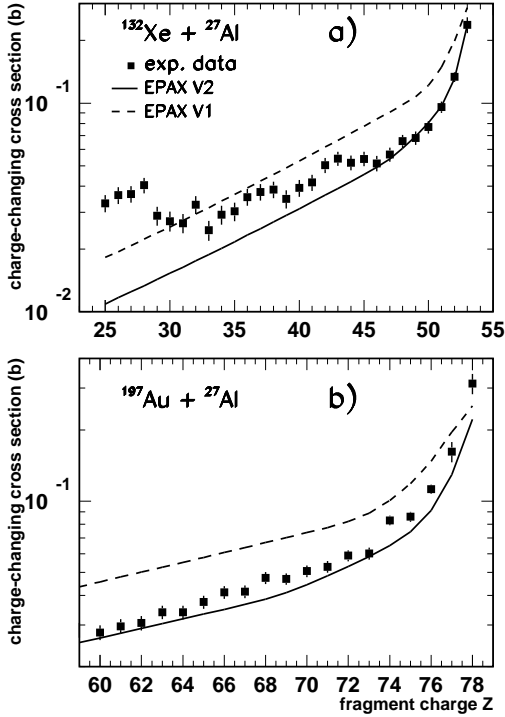,width=0.9\linewidth}
\end{center}
\caption{Experimental charge-changing cross sections
         for $^{132}$Xe (a) and $^{197}$Au ions (b) on $^{27}$Al targets
	 in comparison with
         the old EPAX parametrization Version 1 (dashed line) and the
         new EPAX formula Version 2 (solid line). 
         The experimental data 
         are from Binns {\it et al.}~\protect\cite{binns}.
} 
\label{charge_yield}
\end{figure}

Another comparison of integrated cross sections from {\sc EPAX} to experimental 
data can be performed for charge-changing cross sections. For this purpose, 
we use experimental data from Binns {\it et al.}~\cite{binns}. 
Fig.~\ref{charge_yield} shows a comparison of the experimental data to the 
{\sc EPAX} Versions~1 and~2. As can be seen, nice 
agreement is achieved with the 
new {\sc EPAX} formula, in particular close to the target. 
For the $^{132}$Xe data, the excess charge-changing cross 
sections for lower $Z$ are most likely due to secondary reactions in the
targets which are unsufficiently accounted for.

\subsubsection{Parameters of the residue corridor}

\begin{figure}[bth]
\begin{center}
\epsfig{file=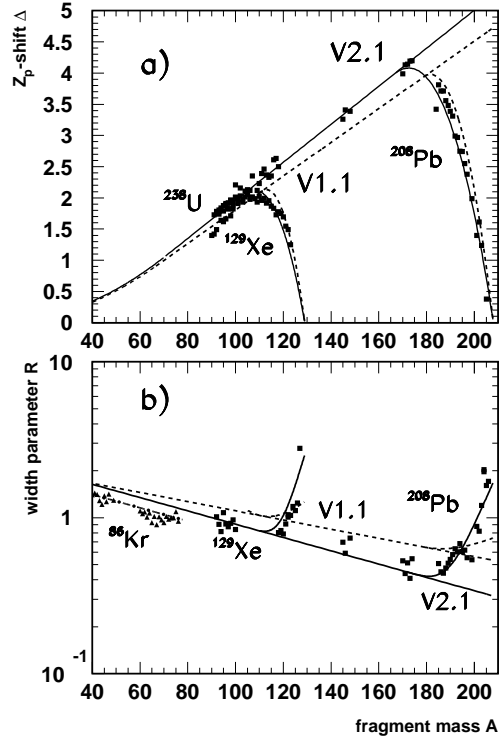,width=0.9\linewidth}
\end{center}
\caption{a) Fragment-mass dependence of the offset, $\Delta$, 
         of $Z_{prob}$ from the line of $\beta$-stability according to
         Eq.~\ref{eq5}. The dashed curves denote
         the old EPAX parametrization, while the modified Version 2 
         is indicated by the full curves. The
         downward-curving lines above $A \approx 110$ and $A \approx 180$ 
         illustrate the return of $Z_{prob}$ to $\beta$-stability 
         close to the projectile (Eq.~\ref{eq9}) for $^{129}$Xe and 
         $^{208}$Pb, respectively, as a projectile.
         Data from fits of Eq.~\ref{eq5} to the results from 
         Refs.~\protect\cite{joerg,dejong,jose} are shown by the full squares.
         b) Fragment-mass dependence of the width parameter $R$ 
         according to Eq.~\ref{eq8}.  The upward-sloping dashed (full) curves
         indicate the shrinking of the widths near the 
         $^{129}$Xe and $^{208}$Pb projectile, respectively,
         for the old (new) version of EPAX according to
         Eq.~\ref{eq10} (Eq.~\ref{eq13}). Results from fits to the data from
         $^{129}$Xe and $^{208}$Pb fragmentation are shown by the full squares;
         those from $^{86}$Kr fragmentation
         (triangles, Ref.~\protect\cite{weber}) have been approximated by
         Eq.~\ref{eq16} (dash-dotted line).
} 
\label{fig_delta_r}
\end{figure}

In the upper part of Fig.~\ref{fig_delta_r}, we show, as a function
of the fragment mass, $A$, the new position $Z_{prob}(A)$
of the ``residue corridor'', for $^{129}$Xe and $^{208}$Pb, 
fragmentation.
Data from $^{238}$U fragmentation (Ref.~\cite{u238}) 
are also included in the Figure,
they were not used to fit the parameters $\Delta_i$ and can serve as a check.
For clarity, we plot the offset, $\Delta$, from 
$Z_{\beta}$ according to Eq.~\ref{eq5}. The dashed lines
are for {\sc EPAX} Version 1;
the full lines represent the new Version 2.
Note that for better results close to the projectile, we have added to
$Z_{prob}$ according to 
Eq.~\ref{eq5} a tiny fragment-mass dependent correction
amounting to $0.002 \cdot A$.
The lines sloping downward above $A \approx 110$ and 
$A \approx 180$ illustrate
the return of $Z_{prob}$ to $\beta$-stability close to the
$^{129}$Xe and $^{208}$Pb projectiles according to Eq.~\ref{eq9}.

In a similar way, the lower part of Fig.~\ref{fig_delta_r}
visualizes the analytical description of the width parameter, $R$.
While the functional form of the correction for $\Delta$,
Eq.~\ref{eq9}, was kept identical with {\sc EPAX} Version 1, the correction
for $R$ near the projectile is now written as 
\begin{equation}
R  =  R \cdot [1+r_1 \cdot A_p \cdot (A/A_p-r_2)^4] \label{eq13}
\end{equation}
for $A/A_p \ge r_2$,
which yields more narrow charge distributions for $A$ close to $A_p$.

Only minor changes were introduced for the slope constants,
$U_n$ and $U_p$, of the neutron and proton-rich sides of the
charge distributions, respectively.
While $U_n$ was increased from 1.5 to 1.65 (constant for all $A$),
$U_p$ is now slightly fragment-mass dependent with
\begin{equation}
U_p  =  U_1 + U_2 \cdot A + U_3 \cdot A^2 \label{eq14}
\end{equation}
which approaches $U_p = 2.2$ for large $A$, compared to a
previous constant value of 2.0 (Gaussian curve).

A close inspection of the width parameter, $R$, 
resulting from fits to the measured $^{86}$Kr
isotope distributions~\cite{weber} shows that they
are slightly but systematically decreased compared to Eq.~\ref{eq8}
(see Fig.~\ref{fig_delta_r}). This means that
the isobar distributions are slightly wider for a neutron-rich
projectile like $^{86}$Kr than for a $\beta$-stable one.
Therefore, we tentatively introduce
the following neutron-excess dependence of $R$:
\begin{equation}
R'  =  R \cdot [1 - 0.0833 \cdot ( Z_{\beta p} - Z_p ) \label{eq16}
\end{equation}
where $R$ is calculated according to Eq.~\ref{eq8}.
For neutron-deficient projectiles, $R$ is not changed.

\begin{figure}[tb]
\vspace{-1cm}
\begin{center}
\epsfig{file=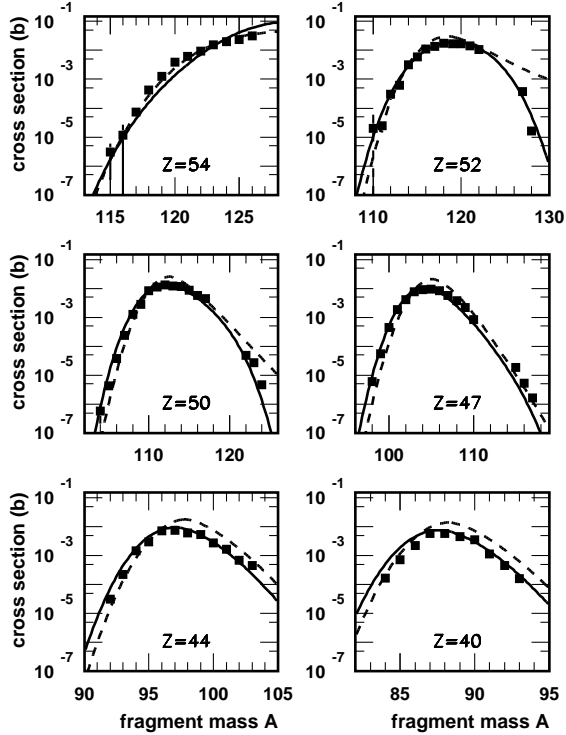,width=1.1\linewidth}
\end{center}
\caption{Isotope distributions from 800 $A$ MeV $^{129}$Xe 
         fragmentation in a $^{27}$Al target~\protect\cite{joerg}
         in comparison with the old EPAX parametrization Version 1
         (dashed curve) and the modified Version 2 (full curve).
} 
\label{fig_xe129sn}
\end{figure}

For a typical example, the combined effect 
of the changes described above is visualized
in Fig.~\ref{fig_xe129sn}, which displays the
isotope distributions from 800 $A$ MeV $^{129}$Xe fragmentation
measured over four orders of magnitude~\cite{joerg}.
In particular, the data measured near the maximum are now much better
reproduced. The stronger increase in $R$ near the projectile mass,
$A_p=129$, leads to a much faster decrease of the corresponding
fragment yields at $A \approx 124$.
Similar observations, though over a smaller vertical range, can be 
made for the fragment distributions from $^{208}$Pb
fragmentation shown in Fig.~\ref{fig_pb}.
It is interesting to note that even for $^{238}$U fragmentation
products in the mass range $100 \lesssim A \lesssim 130$, produced
in violent collisions and correspondingly long chains of statistical
evaporation, rather good agreement between experimental data~\cite{u238}
and EPAX can be observed.

\subsection{Fragmentation of neutron- oder proton-rich projectiles}

\begin{figure}[tb]
\begin{center}
\epsfig{file=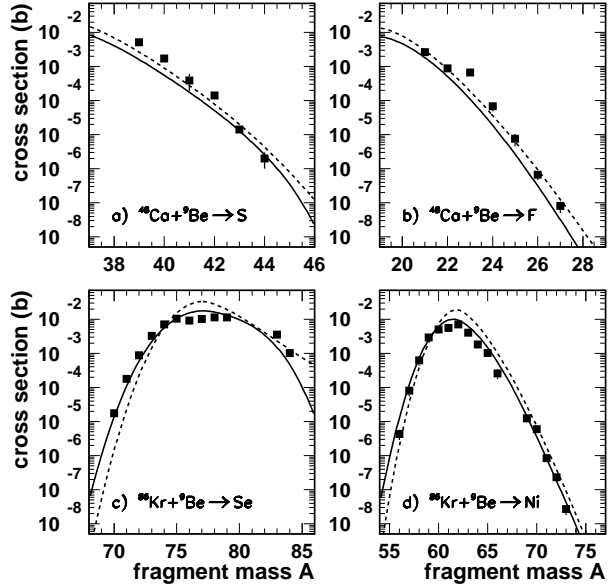,width=1.0\linewidth}
\caption{Examples visualizing the ``memory effect'' for
         neutron-rich projectiles impinging on $^9$Be targets.
         Panels a) and b): S and F isotope distributions from
         $^{48}$Ca fragmentation; c) and d): Se and Ni isotope distributions 
         from $^{86}$Kr fragmentation. 
	 The dashed curves are for EPAX Version 1, 
         the full ones for the present Version 2.
         Experimental data are from Refs.~\protect\cite{ca48,weber}.
} 
\label{fig_kr86_ca48}
\end{center}
\end{figure}

In the previous {\sc EPAX} version~\cite{epaxv11}, 
the parametrization
of the ``memory effect'' was fitted, for the case of
neutron-rich projectiles, to the results from
$^{48}$Ca fragmentation at 213 $A$ MeV~\cite{ca48}
(see Eq.~\ref{eq12}).
For the case of $^{86}$Kr, we were able to check the validity
of this prescription in a different fragment-mass range~\cite{weber}.
We have kept this prescription for the current Version 2,
since the agreement with experimental data has not changed significantly
(see Fig.~\ref{fig_kr86_ca48} for selected examples).

\begin{figure}[tb]
\begin{center}
\epsfig{file=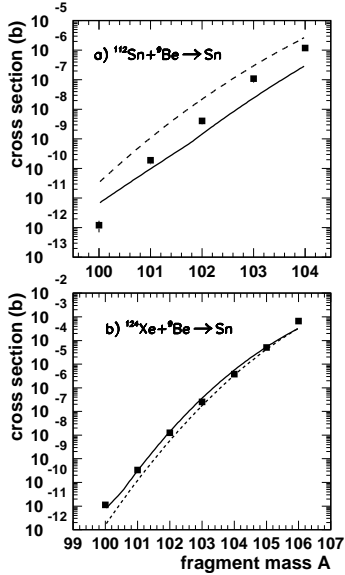,width=0.6\linewidth}
\caption{Experimental isotope distributions for Sn isotopes from
         neutron-deficient projectiles on $^9$Be targets:
         a) from $^{112}$Sn fragmentation~\protect\cite{sn112},
         b) from $^{124}$Xe fragmentation~\protect\cite{xe124}, plotted
         in comparison with the old EPAX parametrization 
         (dashed curves) and the modified version (full curves).
} 
\label{fig_sn}
\end{center}
\end{figure}

The situation is different, however, for the case of
neutron-deficient projectiles.
There, the previous version was unsatifactory from the beginning:
Firstly, there were only very few radiochemical cross sections available
to derive a guess for the memory effect on the 
neutron-deficient side.
Secondly, the resulting formula,
$\Delta_m=[p_2\cdot (A/A_t)^4]\cdot (Z_p-Z_{\beta p})$,
does not yield the limiting value $\Delta_m/(Z_p-Z_{\beta p})=1$ for 
$Z_p=Z_{\beta p}$.
These deficiencies became particularly obvious 
when Schneider {\it et al.}~\cite{sn112}
measured only small cross sections for neutron-removal products from
$^{112}$Sn. This lead to a different functional form
for the memory effect for neutron-deficient projectiles.
The new parametrization reads
\begin{equation}
\Delta_m  = exp\;[p_1 + p_2 \cdot A/A_p] \cdot (Z_p-Z_{\beta p}) \label{eq17}
\end{equation}
which is equivalent to a stronger but more rapidly decaying memory
effect compared to the polynomial (Eq.~\ref{eq12}) used previously.
The consequences of the new parametrization of the memory effect
for neutron-deficient projectiles are visualized in Fig.~\ref{fig_sn}.
The small measured cross sections for $^{100}$Sn~\cite{sn112} are now 
better reproduced,
while the good agreement (over more than seven orders 
of magnitude!) with data measured in
$^{124}$Xe fragmentation~\cite{xe124} is maintained.

We note in passing that Eq.~\ref{eq17} is very similar to what has been
suggested by von Egidy and Schmidt~\cite{egidy} to reproduce
production cross sections from antiproton annihilation. 
The overall agreement with the experimental fragmentation yields
studied in our work
is worse, however, when we adopt their description of the memory effect,
in particular on the neutron-rich side.
Moreover, we do not fit any set of parameters separately for
a specific projectile as has been done in Ref.~\cite{egidy}, 
but rather want to describe all systems
with the same parameter set.

The yield distributions from the neutron-deficient projectile 
$^{58}$Ni, studied in great detail by Blank {\it et al.}~\cite{blank},
deserve special attention. Not only do they represent a rare example
of cross-section measurements down to the sub-nanobarn level,
but also a case where severe discrepancies with {\sc EPAX} Version 1 were
observed: the yields of even-$Z$ isotopes close to the proton drip line
where found to be enhanced in experiment by factors of up to 750~\cite{blank}. 
As a remedy for this deficiency, we have chosen to
switch over from quasi-Gaussian to exponential slopes of the
charge distribution above a certain gradient of the cross-section distribution.
In order to achieve a steady transition between the two slopes, the transition
point and the slope of the exponential have to be adjusted carefully.
For this purpose, we calculate the derivative of 
the logarithm of the cross section 
(Eq.~\ref{eq1}):

\begin{equation}
\frac{dF}{dZ} = \frac{d(log(\sigma))}{dZ} \approx \frac{-2 \cdot R}{ln(10)} 
\cdot (Z-Z_{prob})
\label{eq18}
\end{equation}

The transition point to the exponential slope,
$Z_{exp}$, can be calculated for the proton-rich side
as a function of the
fragment mass $A$ according to 

\begin{equation}
Z_{exp}(A) = Z_{prob}(A) + {\frac{dF}{dZ}\Bigg\vert}_A  \cdot \frac{ln(10)}
{2 \cdot R(A)}
\label{eq19}
\end{equation}

From $Z_{exp}$ on, the slope is exponential with the same gradient as
Eq.~\ref{eq1} at this point.

The gradient for which we switch to the exponential slope
has to be para\-metri\-zed as a function of the fragment mass $A$.
As for the moment only the $^{58}$Ni data exhibit this exponential
trend (the measured cross sections in the $^{100}$Sn region reach the
same cross section level, but are further away from the 
proton drip line), we tried to adjust the function $dF/dZ$ in such a way to 
reproduce the $^{58}$Ni data without deteriorating the good agreement with
the measured $^{124}$Xe
data. The function which fulfills reasonably well these criteria
is the following:

\begin{equation}
\frac{dF}{dZ} = 1.2 + 0.647 \cdot (A/2)^{0.3}
\label{eq20}
\end{equation} 

\begin{figure}[htb]
\begin{center}
\epsfig{file=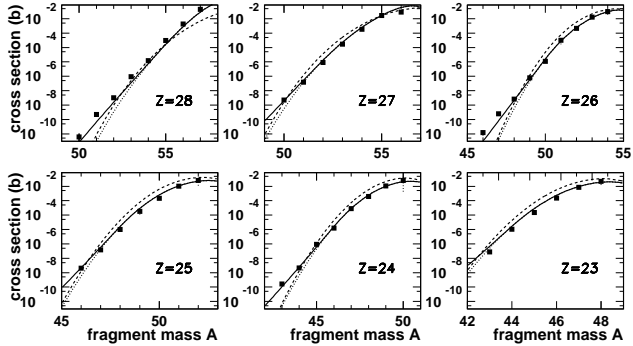,width=1.0\linewidth}
\caption{Experimental isotope distributions for Ni to V isotopes from
         $^{58}$Ni+$^9$Be fragmentation~\protect\cite{blank}
         in comparison with the old EPAX parametrization
         (dashed curves) and the new version (full curves).
         The new EPAX formula without a transition to the exponential slope 
         is shown by the dotted curve.
         This effect is clearly visible below the 10~nb level.
} 
\label{fig_ni58}
\end{center}
\end{figure}

The result for the $^{58}$Ni fragment-yield distributions
is visualized in Fig.~\ref{fig_ni58}.
The full line shows the new {\sc EPAX}
cross sections including the above modification of the slopes, whereas
the omission of the exponential slope leads to the dotted cross-section curve.
As a consequence of this modification, {\sc EPAX} Version 2 predicts a 
production
cross section e.g. for the doubly-magic nucleus $^{48}$Ni 
of $4 \cdot 10^{-13}$ b, which is within reach of
an experiment.

As stated earlier, the present adjustment for very proton-rich projectile 
fragments is only based on results from $^{58}$Ni 
fragmentation~\cite{blank}. Therefore,
caution is advisable when applying this parametrization especially 
to very light proton-rich fragmentation products. It would be interesting
to compare the present parametrization to fragmention yields
at the drip line from other proton-rich beams, e.g. from $^{36}$Ar or $^{78}$Kr.
In the $^{100}$Sn region, the present transition to an 
exponential slope has only a slight influence on the $^{100}$Sn 
cross-section prediction from $^{112}$Sn fragmentation 
($6.6 \cdot 10^{-12}$ b, to be compared to an experimental value
of $1.2 \cdot 10^{-12}$ b~\cite{sn112}).  
It might be interesting to measure fragmentation cross
sections for even more proton-rich nuclei in this
mass region to compare to the modified
{\sc EPAX} formula.

\subsection{Overall quality of {\sc EPAX} Version 2}

Besides the comparison between experimental data and {\sc EPAX} predictions
for individual elemental distributions, we tried also to visualize
in a more global manner the overall quality of the 
new {\sc EPAX} parametrization compared to the data
and the improvements with respect to {\sc EPAX} Version~1.

\begin{figure}[tb]
\begin{center}
\epsfig{file=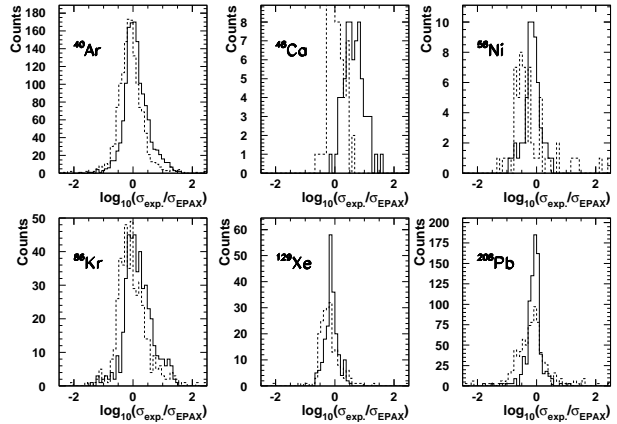,width=1.0\linewidth}
\caption{Logarithm of the ratio between experimental fragmentation cross 
         sections and 
         predictions of EPAX Version 1 (dashed line) and the new 
         EPAX Version 2 (solid line) for projectiles ranging from $^{40}$Ar to 
         $^{208}$Pb. References to the experimental data are given in the
	 text.} 
\label{quality}
\end{center}
\end{figure}

For this purpose, we plot the cross section ratios $\sigma_{exp} / 
\sigma_{EPAX}$ for all projectile-target combinations used in the 
present paper in Fig.~\ref{quality}.
For most of the projectiles the logarithm of the cross-section ratio is 
centered around zero indicating an agreement on average between experiment
and prediction. For the new {\sc EPAX} Version 2, the sigma widths
of these distributions
vary between 0.5 (in the case of $^{40}$Ar and $^{86}$Kr) 
and 0.2 (in the case of $^{129}$Xe). If we analyse all experimental data
together, we obtain a sigma of 0.4. This 
demonstrates that the new {\sc EPAX} formula can predict
cross-sections in most cases within a factor of two.
In almost all cases we observe a significantly 
smaller sigma for the new {\sc EPAX} 
parametrization than for the old version. This is particularly 
striking in the cases of $^{58}$Ni, $^{129}$Xe, and $^{208}$Pb.
Only in the case of $^{48}$Ca the agreement between experimental data 
and predictions deteriorates slightly, exhibiting now a shift to smaller
predicted cross sections. It is likely that the memory effect for
neutron-rich projectiles is the origin of this discrepancy.

\subsection{Comparison with other empirical parametrizations}

As has been mentioned in the Introduction and discussed in detail
in Ref.~\cite{epaxv11}, other empirical parametrizations have been
proposed, mainly for proton-induced spallation reactions
(e.g. Refs.~\cite{st2,webber2}). Contrary to our formula,
these parametrizations fit also the energy dependence
of the cross sections.
The former approach has been extended to describe
heavy-ion induced spallation reactions
by scaling proton-induced cross sections by an energy-dependent
factor~\cite{st3} and
has achieved good agreement with measured cross sections
(in the 100 mb to 1 mb range) for fragments from
medium-mass nuclei (i.e. for projectiles
up to $^{56}$Fe, Ref.~\cite{st3}).

\begin{figure}[htb]
\begin{center}
\epsfig{file=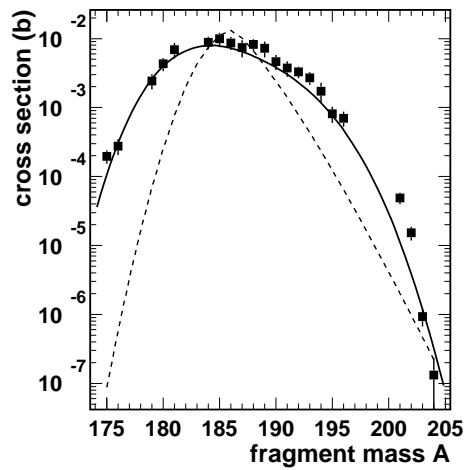,width=0.80\linewidth}
\caption{
Platinum ($Z$=78) isotope distribution from 1 $A$ GeV $^{208}$Pb 
fragmentation in a $^{nat}$Cu target~\protect\cite{dejong}
in comparison with the new EPAX 
parametrization (full curve) and the Tsao {\it et al.}
formula (Ref.~\protect\cite{st3}, dashed curve).
}
\label{fig_silv_tsao}
\end{center}
\end{figure}

For fragmentation cross sections from heavy-mass nuclei,
however, the Tsao {\it et al.} formula~\cite{st3} is less successful.
This can be demonstrated e.g. by
comparing in Fig.~\ref{fig_silv_tsao} their
prediction for the Pt isotope
distribution in the reaction $^{208}$Pb + $^{nat}$Cu
to experimental data~\cite{dejong}.
As we have mentioned in the Introduction, our physical
understanding of high-energy heavy-ion reactions suggests that
proton-induced reactions produce more neutron-rich
isotope distributions of heavy elements than
heavy-ion-induced reactions,
therefore it is unlikely that even a better fit of the Tsao {\it et al.} 
formula to the
data in Fig.~\ref{fig_silv_tsao} would apply also to the 
same isotope distribution formed in the p+$^{208}$Pb reaction.
We believe that separate parameter sets have to be fitted to
the respective experimental data.
For $\approx$1 $A$ GeV protons impinging on heavy targets,
the recent work of Refs.~\cite{farget,enqvist}
provides for the first time a comprehensive data set that allows
to extend previous work~\cite{st2,webber2} to heavier fragments.
A complete parametrization of the bombarding-energy dependence, however,
has to await more measurements at different energies for the heavy systems.

\subsection{{\sc EPAX} predicitions for projectiles very far from stability}

The parameters of the new {\sc EPAX} formula have been adjusted by fitting data
from stable projectiles. As explained above, one of the most difficult
tasks of the present work was to reasonably parametrize the 
cross sections for
fragments close to the projectile. If the projectile is very far
from the line of $\beta$ stability, this task is even more difficult.

The comparisons between {\sc EPAX} predictions and experimental data shown in
the present paper indicate
that the chosen parametrization is reasonable. A possible check 
of its predictive power could be to
compare the {\sc EPAX} formula to other data from neutron-rich or
neutron-deficient projectiles. However, only very few other high-quality
data are available. 

\begin{figure}[tb]
\begin{center}
\epsfig{file=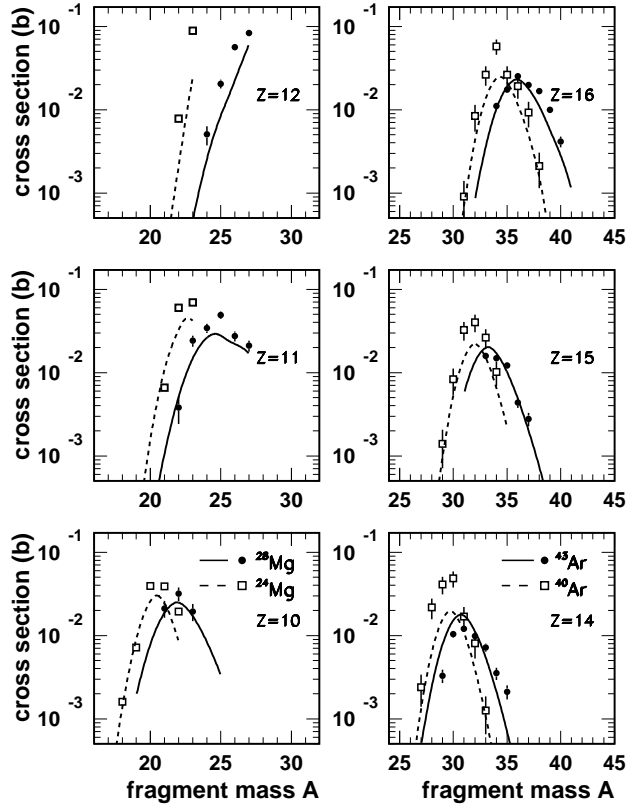,width=1.0\linewidth}
\caption{Experimental cross sections for the fragmentation of stable 
        (open symbols) and radioactive beams (full symbols) compared to 
        the new EPAX parametrization (dashed and full lines, respectively). 
        On the left-hand side, results from $^{24}$Mg~\protect\cite{webber1}
        and $^{28}$Mg~\protect\cite{wan} fragmentation are shown, whereas on the 
        right-hand side $^{40}$Ar~\protect\cite{viyogi}
        fragmentation is compared to $^{43}$Ar~\protect\cite{wan} 
fragmentation.} 
\label{exotics}
\end{center}
\end{figure}

As one check, we compare in Fig.~\ref{exotics} the fragmentation of
a stable $^{24}$Mg beam~\cite{webber1} to a radioactive $^{28}$Mg 
beam~\cite{wan} and of a stable $^{40}$Ar beam~\cite{viyogi} to 
a radioactive $^{43}$Ar beam~\cite{wan}. The elemental
distributions for Mg, Na, and Ne isotopes shown in the left-hand column 
and those for
S, P, and Si isotopes shown in the right-hand column
nicely show the experimental
memory effect and the high quality of the {\sc EPAX} predictions.
In a similar way we compared the results from fragmentation of 
$^{96}$Ru and $^{96}$Zr~\cite{gernhaeuser} to {\sc EPAX} predictions
and found reasonable agreement. However, for projectile beams
still further away from
stability, the memory effect becomes more and
more important. Therefore, we think that some caution should be applied
when using {\sc EPAX} predictions for projectiles very far from stability
like e.g. $^{132}$Sn. Here, no experimental data at all exist to verify
the {\sc EPAX} predictions.

\section{Summary and Conclusions}

We have demonstrated that the quality with which {\sc EPAX} reproduces
measured high-energy fragmentation cross sections could be
improved considerably by introducing rather small modifications to
the formula.
This makes {\sc EPAX} a more reliable tool to predict production rates
for secondary-beam experiments with medium- or heavy-mass
exotic nuclei.

Even after introducing these improvements,
there are still some discrepancies with measured data
which deserve attention in future modifications of the
{\sc EPAX} formula.
One aspect concerns the odd-even effects which were
taken into account by von Egidy and Schmidt~\cite{egidy}. 
At present, we do not think that such a rather small modification (of the
order of 20-30\%) is
necessary in view of the overall discrepancies 
of factors around 2 still observed,
but it may become necessary once better precision can be achieved.
Another open question is the change in slope observed for very
neutron-deficient $^{58}$Ni fragmentation products.
Here it would be desirable to measure systematically the formation
cross sections for fragments close to the proton drip line
for other medium-mass neutron-deficient projectiles, e.g. $^{36}$Ar and 
$^{78}$Kr.

Up to now, we did not use medium-energy data (from reactions at
less then $\approx$100~$A$ MeV) 
which are available from GANIL, RIKEN, or MSU. First of all, these data 
probably do not
fulfill the condition of "limiting fragmentation" mentioned above. In fact, 
it is well known that at these energies nucleon-exchange reactions as well
as pick-up reactions become increasingly important and thus alter in particular
the cross sections close to the projectile. A striking example for this 
is the difference in experimental cross sections for $^{100}$Sn production
from a $^{112}$Sn beam of roughly a factor of 100 in recent 
GANIL and GSI experiments~\cite{xe124,sn100}. 
In addition, these
intermediate-energy data are measured with separators having very low
transmissions (of the order of few percent) which are difficult to
measure experimentally.
It would be interesting, however, to compare {\sc EPAX} also to cross sections
obtained at lower incident energies 
once high-quality data become available.

A FORTRAN program of the {\sc EPAX} formula
as well as additional graphs comparing experimental data to {\sc EPAX}
predictions can be downloaded from
the following address: ~ftp://ftpcenbg.in2p3.fr/pub/nex/epax.

\section*{Acknowledgments}

We wish to thank our colleagues Jose Benlliure, 
Roman Gernh\"auser, Manuel de~Jong, Akira Ozawa, 
J\"org Reinhold, Robert Schneider, Shuying Wan, and Martin Weber 
for providing their 
experimental cross sections prior to publication.
One of us (K.S.) acknowledges the kind hospitality of the 
Centre d'Etudes Nucl\'eaires at Bordeaux-Gradignan where part of
this work was completed.

\end{document}